\documentstyle[12pt,epsf]{article}

\newcommand{\beq}{\begin{equation}}
\newcommand{\eeq}{\end{equation}}
\newcommand{\bea}{\begin{eqnarray}}
\newcommand{\eea}{\end{eqnarray}}
\newcommand{\rar}{\rightarrow}

\newcommand{\lan}{\langle}
\newcommand{\ran}{\rangle}

\begin{document}

\font\fortssbx=cmssbx10 scaled \magstep2
\hbox to \hsize{
\includegraphics{uwlogo.ps}
\hskip.5in \raise.1in\hbox{\fortssbx University of Wisconsin - Madison}
\hfill$\vcenter{\hbox{\bf MADPH-95-900}
            \hbox{August 1995}}$ }
\vskip 2cm
\begin{center}
\Large
{\bf Radiative rare $B$ decays revisited} \\
\vskip 0.5cm
\large
  Sini\v{s}a Veseli  and M. G. Olsson  \\
\vskip 0.1cm
{\small \em Department of Physics, University of Wisconsin, Madison,
	\rm WI 53706}
\end{center}
\thispagestyle{empty}
\vskip 0.7cm

\begin{abstract}
We reexamine contributions of higher $K$-resonances to the
radiative rare decays $b\rar s\gamma$ in the limit where both
$b$- and $s$-quark are considered heavy. Using the non-relativistic
quark model, and the form factor definitions
consistent with the HQET covariant trace formalism, we
 find significant disagreement with previous work which also used
heavy quark symmetry, and
 excellent agreement with experimental results. In particular,
the two largest fractions of the inclusive $b\rar s\gamma$
branching ratio are found to be
$(16.8\pm 6.4)\%$ for $B\rar K^{*}(892)\gamma$ and
$(6.2\pm 2.9)\%$
 for $B\rar K_{2}^{*}(1430)\gamma$ decays. We also compare
the contribution from the radiative  decays into the eight
$K$-meson states to the
inclusive experimental $b\rar s\gamma$ mass
distribution.
\end{abstract}

\newpage

\section{Introduction}

Flavor changing neutral current transitions involving the $B$-meson
provide a unique opportunity to study the electroweak theory in
higher
orders. Although transitions like $b\rar s\gamma$, $b\rar se^{+}e^{-}$,
and $b\rar sg$ vanish at the tree level, they can be described
by one loop (``penguin'') diagrams, in which a $W^{-}$ is emitted and
reabsorbed \cite{shifman}. These processes occur at a rate small enough
to be sensitive to  physics beyond the Standard Model
\cite{hewett}. Similar
flavor violating processes in the $K$-meson system have the disadvantage
that non-perturbative long distance effects are quite large, and
it is difficult to extract the quark level physics from
well known processes like $K^{+}\rar \pi^{+}e^{+}e^{-}$.

Among all rare $B$ decays, radiative processes
$B\rar X_{s}\gamma$ (especially decay
$B\rar K^{*}(892)\gamma$) have received an increasing attention,
because of the experimental measurement of the $B\rar K^{*}(892)\gamma$
exclusive branching ratio \cite{ammar},
\beq
BR(B\rar  K^{*}(892)\gamma) = (4.5\pm 1.5\pm 0.9)\times 10^{-5}\ .
\eeq
which has been
recently updated \cite{cleoex} to
\beq
BR(B\rar  K^{*}(892)\gamma) = (4.3^{+1.1}_{-1.0}\pm 0.6)\times 10^{-5}\ ,
\eeq
and also of the inclusive rate \cite{alam},
\beq
BR(B\rar  X_{s}\gamma) = (2.32\pm 0.57\pm 0.35)\times 10^{-4}\ .
\eeq
Several methods have been employed to predict
exclusive $B\rar K^{*}(892)\gamma $ decay rate: HQET \cite{ali2,mannel},
QCD sum rules \cite{dominguez}-\cite{narison}, quark models
\cite{donnell}-\cite{tang}, bound state resonances
\cite{atwood}, and Lattice QCD
\cite{bernard}-\cite{burford}. The theoretical uncertainty, which
was originally of two orders of magnitude, has been greatly reduced
in the more recent studies. However, there is still a large
spread between different results.

In this paper we follow the approach of \cite{ali2,mannel}, in
which both
$b$- and $s$-quark are considered heavy. In the heavy quark limit
the long distance effects are contained within   unknown
form factors, whose precise definition consistent with the covariant
trace formalism \cite{georgi2}-\cite{falk}
has been clarified only recently \cite{modelling}.
This is precisely
the reason why our results substantially
 differ from \cite{mannel}, even though we use the
same non-relativistic quark model for the wave functions of the light
degrees of freedom (LDF).
Our results show that the
 ratio of the exclusive
 $B\rar K^{**}\gamma$ to the inclusive decay rate $B\rar X_{s}\gamma$
 was underestimated
for the channel $B\rar K^{*}(892) \gamma$
($(16.8\pm 6.4)\%$ as opposed to $(3.5-12.2)\%$ from \cite{mannel}),
and significantly overestimated for the decay
$B\rar K_{2}^{*}(1430) \gamma$ ($(6.2\pm 2.9)\%$ as
opposed to $(17.3-37.1)\%$ from \cite{mannel}). We  emphasize
that our prediction for the decay $B\rar K^{*}(892) \gamma$
is in agreement with experimental result of $(19\pm 5)\%$.
Although other exclusive decays have
not yet been identified, we have compared with experiment
the contribution
from the eight $B\rar K^{**}\gamma$ decays to the inclusive
$B\rar X_{s}\gamma$ mass distribution.

The paper is organized as follows: in Section \ref{th}
we restate the theoretical framework for the  $B\rar K^{**}\gamma$
decays. Section \ref{ff} contains a discussion of the form factor
calculation. The expressions for the form factors given in
\cite{modelling} are evaluated in terms of the wave functions and
energies
of the light degrees of freedom  in the meson rest frame.
We discuss here the model used in establishing
the LDF wave functions and energies. An extensive literature
exists in this subject, so we have attempted to set our results
in context with previous
calculations in Section \ref{res}. Our conclusions are summarized
in Section \ref{con}.

\section{Theory of $B\rar K^{**}\gamma$ decays}
\label{th}

The effective Hamiltonian for the decays $B\rar X_{s}\gamma$ can be
found in many places, e.g. \cite{grinstein}-\cite{buras}. It is derived
by integrating out the top quark and $W$-boson at the same scale
$\mu\approx M_{W}$. An appropriate operator basis for the
effective Hamiltonian consists
of four-quark operators and the magnetic moment type operators
of dimension six (${\cal O}_{1}-{\cal O}_{8}$). Higher dimensional
operators are suppressed by powers of the masses of the heavy particles.
For the $B\rar K^{**}\gamma$ decays only the operator
${\cal O}_{7}$ contributes, so that
\beq
H_{eff} = -\frac{4G_{F}}{\sqrt{2}} V_{tb} V^{*}_{ts} C_{7}(m_{b})
{\cal O}_{7}(m_{b})\ .
\eeq
Here,
${\cal O}_{7}$ is given by
\beq
{\cal O}_{7} = \frac{e}{32\pi^{2}}F_{\mu\nu} [
m_{b}\bar{s}\sigma^{\mu\nu}(1+\gamma_{5})b
+m_{s}\bar{s}\sigma^{\mu\nu}(1-\gamma_{5})b]\ ,\label{o7}
\eeq
with $\sigma^{\mu\nu}=\frac{i}{2}[\gamma^{\mu},\gamma^{\nu}]$.
The explicit expression
for the Wilson coefficient $C_{7}(m_{b})$ as a function of
$\frac{m_{t}^{2}}{M_{W}^{2}}$
can be found in \cite{buras,deshpande}.
The value of $C_{7}$ can be calculated perturbatively
at the mass scale $\mu = M_{W}$.  The evolution from $M_{W}$
down to a mass scale $\mu=m_{b}$ introduces large QCD corrections.
This proceedure also introduces large theoretical uncertainties, primarily
due to the choice of the renormalization scale $\mu$ (taken above
as $m_{b}$), which can be as large as $25\%$ \cite{buras}.

As proposed in \cite{ali2, mannel}, we evaluate
the hadronic matrix element of ${\cal O}_{7}$ between a $B$-meson in
the initial state, and a generic $K^{**}$-meson in the final state,
in the heavy quark limit for the $b$- and $s$-quarks.
Matrix elements of bilinear currents of two heavy quarks
($J(q)=\bar{Q'}\Gamma Q$) are
most conveniently evaluated within the framework of the trace
formalism, which was formulated in \cite{georgi2,korner} and generalized
to  excited states in \cite{falk}. Denoting
$\omega=v\cdot v'$, where $v$ and $v'$ are the four-velocities
of the two mesons mesons, we have
\beq
\lan \Psi'(v')|J(q)|\Psi(v)\ran =  {\rm Tr}[\bar{M'}(v')\Gamma M(v)]
{\cal M}(\omega)\ ,\label{tr}
\eeq
where $M'$ and $M$ denote matrices describing states $\Psi'(v')$ and
$\Psi(v)$,
$\bar{M}=\gamma^{0}M^{\dag}\gamma^{0}$, and
${\cal M}(\omega)$ represents the LDF.
For all transitions considered in this paper matrices $M$ and $M'$, as well
as definitions for ${\cal M}(\omega)$, can be found in
\cite{mannel,modelling}.
Using (\ref{tr}), we can write
\beq
\lan K^{**}\gamma | {\cal O}_{7}(m_{b}) | B\ran =
\frac{e}{16\pi^{2}}\eta_{\mu}q_{\nu} {\rm Tr}[\bar{M'}(v')
 \Omega^{\mu\nu} M(v)]{\cal M}(\omega)\ ,
\label{omunu}
\eeq
where the factor $q_{\nu}=m_{B}v_{\nu}- m_{K^{**}}v'_{\nu}$
came from the
derivative in the field strength $F_{\mu\nu}$ of (\ref{o7}), $\eta_{\mu}$ is
the photon polarization vector, and
\beq
\Omega^{\mu\nu} = m_{B}\sigma^{\mu\nu}(1+\gamma_{5})
+
 m_{K^{**}}\sigma^{\mu\nu}(1-\gamma_{5})\ .
\eeq
Expression (\ref{omunu}) can be further simplified
using $\not{v}M(v)=M(v)$.

Now, using the mass shell condition of
the photon ($q^{2}=0$), and polarization sums for spin-1 and spin-2
particles, we obtain the following decay rates \cite{mannel}:
\bea
\Gamma(B\rar K^{*}(892)\gamma) &=& \Omega |\xi_{C}(\omega)|^{2}
\frac{1}{y}[(1-y)^{3}(1+y)^{5}(1+y^{2})]\ ,\label{g1}\\
\Gamma(B\rar K_{1}(1270)\gamma) &=& \Omega |\xi_{E}(\omega)|^{2}
\frac{1}{y}[(1-y)^{5}(1+y)^{3}(1+y^{2})]\ ,\\
\Gamma(B\rar K_{1}(1400)\gamma) &=& \Omega |\xi_{F}(\omega)|^{2}
\frac{1}{24y^{3}}[(1-y)^{5}(1+y)^{7}(1+y^{2})]\ ,\\
\Gamma(B\rar K^{*}_{2}(1430)\gamma) &=& \Omega |\xi_{F}(\omega)|^{2}
\frac{1}{8y^{3}}[(1-y)^{5}(1+y)^{7}(1+y^{2})]\ ,\\
\Gamma(B\rar K^{*}(1680)\gamma) &=& \Omega |\xi_{G}(\omega)|^{2}
\frac{1}{24y^{3}}[(1-y)^{7}(1+y)^{5}(1+y^{2})]\ ,\\
\Gamma(B\rar K_{2}(1580)\gamma) &=& \Omega |\xi_{G}(\omega)|^{2}
\frac{1}{8y^{3}}[(1-y)^{7}(1+y)^{5}(1+y^{2})]\ ,\\
\Gamma(B\rar K^{*}(1410)\gamma) &=& \Omega |\xi_{C_{2}}(\omega)|^{2}
\frac{1}{y}[(1-y)^{3}(1+y)^{5}(1+y^{2})]\ ,\\
\Gamma(B\rar K_{1}(1650)\gamma) &=& \Omega |\xi_{E_{2}}(\omega)|^{2}
\frac{1}{y}[(1-y)^{5}(1+y)^{3}(1+y^{2})]\ ,
\label{g8}
 \eea
where we used abbreviations
\bea
y&=&\frac{m_{K^{**}}}{m_{B}}\ ,\\
\Omega &=& \frac{\alpha}{128\pi^{4}}G_{F}^{2}m_{b}^{5}|V_{tb}|^{2}
|V_{ts}|^{2}|C_{7}(m_{b})|^{2}\ ,
\eea
and the argument  of the Isgur-Wise (IW) functions is fixed
by the mass shell condition of the photon ($q^{2}=0$),
\beq
\omega = \frac{1+y^{2}}{2y}\ .
\eeq
Note that in the expressions for the decay rates (\ref{g1})-(\ref{g8})
given in \cite{mannel}, a factor of $(1-y^{2})$ was omitted. Also,
as observed in \cite{mannel}, since decays
into the states belonging to the same spin symmetry doublet
are described by the same Isgur-Wise function, and since
in the heavy-quark limit the two members of a spin doublet
are degenerate in mass, from (\ref{g1})-(\ref{g8}) one has
\bea
\Gamma(B\rar K^{*}_{2}(1430)\gamma)
&\approx& 3\Gamma(B\rar K_{1}(1400)\gamma) \ ,\label{rel1}\\
\Gamma(B\rar K_{2}(1580)\gamma)
&\approx& 3\Gamma(B\rar K_{1}(1680)\gamma) \ .\label{rel2}
\eea
As indicated, these relations are only approximate due
to a large breaking of the spin symmetry for the $s$-quark.

\section{Model for the Isgur-Wise functions}
\label{ff}

As already mentioned, even though we use the same
non-relativistic quark model,
our calculation differs significantly from
\cite{mannel}  in evaluation of the IW form factors needed
for the decay rates.
Assuming that we can describe heavy-light mesons using a
non-relativistic potential model, the rest frame LDF wave functions
(with angular momentum $j$ and its projection $\lambda_{j}$),
can be written as
\beq
\phi^{(\alpha L)}_{j\lambda_{j}}({\bf x})= \sum_{m_{L},m_{s}}
R_{\alpha L}(r)Y_{Lm_{L}}(\Omega)\chi_{m_{s}}
\lan L,m_{L};\frac{1}{2},m_{s}|j,\lambda_{j};L,\frac{1}{2}\ran\ ,
\eeq
where $\chi_{m_{s}}$ represent the rest frame spinors normalized
to one, $\chi^{\dag}_{m'_{s}}\chi_{m_{s}}=\delta_{m'_{s},m_{s}}$, and
$\alpha$ represents all other quantum numbers.
According
to \cite{modelling}, instead of the simple
overlap of the two wave functions,
the form factor definitions should include a Lorentz
invariant factor in front of the overlap of the two wave functions
describing the initial and the
final states of the LDF.
Also, following the suggestion of \cite{zalewski2},
overlaps of the two LDF wave functions
can be done in the Breit frame (${\bf v}=-{\bf v}'$), where
 the boost factors (connecting the moving
to rest LDF states) cancel out.
All this leads to the following expressions
\cite{modelling} valid
for the non-relativistic quark model\footnote{As
 pointed out in \cite{modelling}, models based
on the Dirac equation with a central potential lead to the
same expressions for the IW functions.}
 (suppressing
quantum numbers $\alpha'$ and $\alpha$, and using the notation
of \cite{mannel}):
\bea
\xi_{C}(\omega)& =&
\frac{2}{\omega+1}
\lan j_{0}(ar)\ran_{00}
\ ,
\hspace*{+2.65cm}0_{\frac{1}{2}}^{-}\rar (0_{\frac{1}{2}}^{-},
1_{\frac{1}{2}}^{-})\ ,
\label{xic}\\
\xi_{E}(\omega)& =&
\frac{2}{\sqrt{\omega^{2}-1}}
\lan j_{1}(ar)\ran_{10}
 \ ,
\hspace*{+2.1cm}0_{\frac{1}{2}}^{-}\rar (0_{\frac{1}{2}}^{+},
1_{\frac{1}{2}}^{+})\ ,\\
\xi_{F}(\omega)& =&
\sqrt{\frac{3}{\omega^{2}-1}}\frac{2}{\omega+1}
\lan j_{1}(ar)\ran_{10} \ ,
\hspace*{+1cm}
0_{\frac{1}{2}}^{-}\rar (1_{\frac{3}{2}}^{+},
2_{\frac{3}{2}}^{+})\ , \\
\xi_{G}(\omega)& =&
\frac{2\sqrt{3}}{\omega^{2}-1}
\lan j_{2}(ar)\ran_{20} \ ,
\hspace*{+2.4cm}
0_{\frac{1}{2}}^{-}\rar (1_{\frac{3}{2}}^{-},
2_{\frac{3}{2}}^{-})\
\label{xig},
\eea
where (denoting the energy of the LDF as $E_{\bar{q}}$),
\beq
a = (E_{\bar{q}}+E'_{\bar{q}})\sqrt{\frac{\omega-1}{\omega+1}}\ ,
\eeq
and
\beq
\lan F(r) \ran_{L'L}^{\alpha'\alpha} = \int r^{2} dr R^{*}_{\alpha'L'}(r)
R_{\alpha L}(r)F(r)\ .
\eeq
Note that  (\ref{xic})-(\ref{xig})
 include transitions from the ground state
into radially
 excited states. If the two $j=\frac{1}{2}$ states are the same,
$E'_{\bar{q}}=E_{\bar{q}}$ and
$\xi_{C}$ is normalized to one.
The above expressions should be compared with the ones
used in \cite{mannel} (putting a
tilde over the
form factors to avoid confusion),
\bea
\tilde{\xi}_{C}(\omega)& =& \lan j_{0}(\tilde{a}r)\ran_{00}\ ,
\\
\tilde{\xi}_{E}(\omega)& =& \sqrt{3}\lan j_{1}(\tilde{a}r)\ran_{10}\ ,
\\
\tilde{\xi}_{F}(\omega)& =& \sqrt{3}\lan j_{1}(\tilde{a}r)\ran_{10}\ ,
\\
\tilde{\xi}_{G}(\omega)& =& \sqrt{5}\lan j_{2}(\tilde{a}r)\ran_{20}\ ,
\eea
with the definition
\beq
\tilde{a}=E'_{\bar{q}}\sqrt{\omega^{2}-1}\ .
\eeq

For the numerical estimates we employ the
model used in \cite{isgw} (usually referred to as
the ISGW model), the Schr$\ddot{\rm o}$dinger equation with
\beq
V(r) = -\frac{4\alpha_{s}}{3r} + c + b r\ .
\eeq
With sensible choice of parameters,
this simple model gives quite reasonable
spin-averaged spectra of $b\bar{d}$ and $s\bar{d}$ mesons up to $L=2$.
However, instead of just using a single harmonic oscillator
wave function (as was done in \cite{mannel}), for the
radial wave function
of the LDF, we
numerically solve the Schr$\ddot{\rm o}$dinger equation.
To determine the parameters of the model, we fix
$b=0.18\ GeV^{2}$ (which was also used in \cite{isgw}), and vary
$\alpha_{s}$ and $c$ for a given value of $m_{u,d}$
(in the range $0.30-0.35\ GeV$), and $m_{s}$ (in the range
$0.5-0.6\ GeV$),
 until  a good description of the spin averaged spectra of $K$-meson states
is obtained.
Following this proceedure,
our $\alpha_{s}$ ranges from $0.37$ to $0.48$, while $c$ takes values
from $-0.83\ GeV$ to $-0.90\ GeV$. These parameters are in
good agreement with the original ISGW values \cite{isgw}
($\alpha_{s}=0.50$ and $c=-0.84\ GeV$ for $m_{u,d}=0.33\ GeV$ and
$m_{s}=0.55$). We emphasize that the original ISGW parameters give
results
that are well inside the ranges for all decays quoted in this paper.
By varying
the $c$- and $b$-quark masses we could also obtain good spin averaged
description
of the $B$ and $D$ mesons. However, to be consistent
with heavy quark symmetry, the
wave function  for the $B$ meson was chosen to be
the same as the one obtained for the spin averaged
(ground state for $L=0$)
$K$ and $K^{*}(892)$ mesons.

To completely
define our proceedure, we have to specify how
the LDF energy $E_{\bar{q}}$ was determined.
In \cite{mannel} for a given $K^{**}$-meson the LDF energy
was defined as
\beq
E_{\bar{q}}=\frac{m_{K^{**}}*m_{u,d}}{m_{s}+m_{u,d}}\ .
\eeq
This definition was proposed to  account for
the fact that $s$ mesons aren't particularly
heavy. On the other hand, a definition that is consistent with
 heavy quark symmetry is
\beq
E_{\bar{q}}= m_{K^{**}}-m_{s}\ .
\eeq
It should be noted that these two expressions are not equivalent
in the heavy quark limit.
In order to explore the sensitivity of our results
on the choice of $E_{\bar{q}}$, we have repeated all calculations
employing both of these two definitions, and in the final results
we have quoted the broadest possible
 range obtained for the form factors (and for all other results).
 Finally, $E_{\bar{q}}$ for the $B$ meson
 has been taken to be
 the same as $E_{\bar{q}}$ for the $K^{*}(892)$
meson, consistent with heavy quark symmetry.
 It turns out that this
is actually a very reasonable assumption.
The range of $E_{\bar{q}}$ that was used here for
$B$ and $K^{*}(892)$ meson was from $0.296\ GeV$ to $0.396\ GeV$.
On the other hand,
 from the $CLEO$ data on
the semileptonic $B$ decays \cite{cleo},
 and the LQCD heavy-light wave
function \cite{duncan}, it was estimated \cite{iw} that in $B$
systems $E_{\bar{q}}$ ranges from
$0.266\ GeV$ to $0.346\ GeV$.

We believe that the proceedure outlined above enables us to
estimate a reasonable range for the unknown IW
form factors in a physically more acceptable way than it
was done in \cite{mannel}, by simply varying the scale parameter of
the single harmonic oscillator wave function.

\section{Our results and comparison with previous investigations}
\label{res}

In Table \ref{tab1} we present our results for the range
of (absolute) values of the form factors at the indicated
value of $\omega$, for the ratio $R=\frac{\Gamma(B\rar
K^{**}\gamma)}{\Gamma(B\rar X_{s}\gamma)}$,
and for the branching ratio $BR(B\rar K^{**}\gamma)$, for the
various $K^{**}$-mesons.
The inclusive
branching ratio $B\rar X_{s}\gamma$ is usually taken
to be QCD improved quark decay rate for $b\rar s\gamma$,
which
can be written as \cite{grinstein,buras,deshpande}
\beq
\Gamma(B\rar X_{s}\gamma ) = 4\Omega
(1-\frac{m_{s}^{2}}{m_{b}^{2}})^{3}(1+\frac{m_{s}^{2}}{m_{b}^{2}})\ .
\eeq
The leading log prediction for
$BR(b\rar s\gamma)$ is $(2.8\pm 0.8)\times 10^{-4}$
\cite{buras,deshpande}, where the uncertainty is due to
the choice of the QCD scale. The next-to-leading order terms
that have been calculated tend to reduce the prediction to
about $1.9\times 10^{-4}$ \cite{ciuchini}. Both of these predictions
are in excellent agreement with the recent
experimental result of
$BR(b\rar s\gamma)=(2.32\pm 0.57\pm0.35)\times 10^{-4}$
 \cite{alam}.
For the numerical values of the $B\rar K^{**}\gamma$
branching ratios given in Table \ref{tab1} we used the leading log result
of
$BR(b\rar s\gamma)=2.8\times 10^{-4}$.

In order to make comparison of our results with previous calculations
easier, we have tabulated our results together with results of
\cite{mannel} and \cite{altomari} in Table \ref{tab2}. As far as we
know, these
two papers are the only ones that have dealt with radiative
rare $B$ decays into higher $K$-resonances. There
has been much more work done on the decay $B\rar K^{*}(892)\gamma)$,
and we have tabulated some of these results in Table \ref{tab3}.
As one can see from Table \ref{tab3}, the predictions
for this particular ratio ranges from a $0.7\%$ \cite{hassan} to
$97.0\%$ \cite{donnell}.
The data suggest
a value of $(19\pm 5)\%$. Note that
our result of
$(16.8\pm 6.4)\%$
is consistent with the data,  unlike the values
quoted in  \cite{mannel} and \cite{altomari}. As far as decays
into higher $K$ resonances are concerned, our results are
in general in much better agreement with \cite{altomari} than with
\cite{mannel}. In particular, the authors of \cite{mannel}
emphasized a large branching ratio
for the decay $B\rar K_{2}^{*}(1430)\gamma$ $((17.3-37.1)\%)$,
while our results indicate a 3-6 times smaller value of
$(6.2\pm 2.9)\%$, a result which agrees
with the one quoted in \cite{altomari} ($6.0\%$). Also note that
our numerical results from Table \ref{tab2}
support relations (\ref{rel1}) and (\ref{rel2}).

With the exception of the $K^{*}(892)\gamma$ channel, no other
exclusive radiative processes have been identified so far.
The inclusive radiative $B\rar X_{s}\gamma$ mass distribution
has however been measured by $CLEO$ \cite{alam}, and is shown
in Fig. \ref{fig1}. We have
normalized experimental data
so that the integrated distribution gives unity. The
$K^{*}(892)$ peak is evident, but the higher mass
contribution are not resolved. We have attempted to model this
inclusive distribution by considering the contributions from each of the
exclusive $K^{**}\gamma$ channels considered in this paper
(and given in Table \ref{tab2}).

In order to compare our result to experiment, we  replace a given
$R_{K^{**}}$ by a mass distribution reflecting the finite total width
 $\Gamma_{K^{**}}$
of the $K^{**}$ resonance \cite{pdg},
\beq
\frac{dR(m_{X_{s}})}{dm_{X_{s}}} = \sum_{K^{**}}
\frac{R_{K^{**}}}{\pi} \frac{\Gamma_{K^{**}}/2}{(m_{X_{s}}-
m_{K^{**}})^{2}+(\Gamma_{K^{**}}/2)^{2}}\ .
\label{md}
\eeq
The integrated distribution gives
\beq
\int \frac{dR(m_{X_{s}})}{dm_{X_{s}}} dm_{X_{s}}
= \sum_{K^{**}} R_{K^{**}}\ .
\eeq
In Figure \ref{fig1} we show the total resonance contribution
(solid line) compared to  the experimental inclusive
$B\rar X_{s}\gamma$ mass distribution. The area of the resonance curve
is $37.4\%$ of the total inclusive rate (see Tables \ref{tab1} or
\ref{tab2}). We see the general shape is correct, but it
is difficult to make more quantitative statements due to the large errors
involved.

\section{Conclusion}
\label{con}

In this paper we have reexamined predictions of  heavy quark symmetry
for the radiative rare decays of $B$-mesons into higher $K$-resonances.
An earlier calculation \cite{mannel} suggested a substantial fraction
($(17.3-37.1)\%$) of the inclusive $b\rar s\gamma$  branching
ratio going into the $K_{2}^{*}(1430)$ channel, and only
$(3.5-12.2)\%$ going into the $K^{*}(892)$ channel.
Even though we used the same non-relativistic quark model,
our calculation yields fractions of $(16.8\pm 6.4)\%$ and
$(6.2\pm 2.9)\%$ for $K^{*}(892)$ and $K_{2}^{*}(1430)$ channels,
respectively. Note that experimental
results favor the value of $(19 \pm 5)\%$ for the $K^{*}(892)$
channel.
Besides a more careful treatment of the uncertainty
in the wave functions of the light degrees of freedom,
our calculation differs from \cite{mannel} in employing
 form factor definitions that are consistent with
the HQET covariant trace formalism \cite{modelling}. As a consequence
of that, our results for all decay channels
significantly differ from \cite{mannel}.
The contribution of the eight
$K^{**}\gamma$ channels to the inclusive
$B\rar X_{s}\gamma$ mass distribution was compared with experiment.
We find the general shape of the mass spectrum to
be correct, but due to the large errors involved
one cannot reach more quantitative
conclusions.

\vskip 1cm
\begin{center}
ACKNOWLEDGMENTS
\end{center}
This work was supported in part by the U.S. Department of Energy
under Contract No. DE-FG02-95ER40896 and in part by the University
of Wisconsin Research Committee with funds granted by the Wisconsin Alumni
Research Foundation.

\newpage
\begin{figure}
\begin{center}
TABLES
\end{center}
\end{figure}

\begin{table}
\caption{Our results for the
range of absolute values of the form factors
at indicated value of $\omega$, for the ratio
$R=\frac{\Gamma(B\rar K^{**}\gamma)}{\Gamma(B\rar X_{s}\gamma)}$,
and for the branching ratio $BR(B\rar K^{**}\gamma)$, for the
various $K^{**}$-mesons. For the calculation
of branching ratios we used the value $\Gamma(B\rar X_{s}\gamma)=
2.8\times 10^{-4}$ \protect\cite{buras}. }
\label{tab1}
\begin{center}
\begin{tabular}{|cccccccc|}
\hline
Meson         &     State     &   $J^{P}$  &$j$  & $\omega$&
$\xi$     &   $R[\%]$       &   $BR \times 10^{5}$ \\
\hline
$K$    &  $C$     & $0^{-}$  &$\frac{1}{2}$ & \multicolumn{4}{c|}{forbidden} \\
$K^{*}(892)$ & $C^{*}$ &  $1^{-}$ &$\frac{1}{2}$&
 $3.031$ & $0.289\pm 0.057$ &
$16.8\pm 6.4$ & $4.71\pm 1.79$ \\
$K^{*}(1430)$ & $E$ &  $0^{+}$&$\frac{1}{2}$ &
 \multicolumn{4}{c|}{forbidden} \\
$K_{1}(1270)$ & $E^{*}$ &  $1^{+}$ &$\frac{1}{2}$&
 $2.194$ & $0.277\pm 0.053$ &
$4.3\pm 1.6$ & $1.20\pm 0.44$ \\
$K_{1}(1400)$ & $F$ &  $1^{+}$ &$\frac{3}{2}$&
 $2.016$ & $0.171\pm 0.040$ &
$2.1\pm 0.9$ & $0.58\pm 0.26$ \\
$K^{*}_{2}(1430)$ & $F^{*}$ &  $2^{+}$ &$\frac{3}{2}$&
 $1.987$ & $0.175\pm 0.043$ &
$6.2\pm 2.9$ & $1.73\pm 0.80$ \\
$K^{*}(1680)$ & $G$ &  $1^{-}$ &$\frac{3}{2}$&
 $1.702$ & $0.241\pm 0.035$ &
$0.5\pm 0.2$ & $0.15\pm 0.04$ \\
$K_{2}(1580)$ & $G^{*}$ &  $2^{-}$ &$\frac{3}{2}$&
 $1.820$ & $0.203\pm 0.024$ &
$1.7\pm 0.4$ & $0.46\pm 0.11$ \\
$K(1460)$    &  $C_{2}$     & $0^{-}$  &$\frac{1}{2}$ &
\multicolumn{4}{c|}{forbidden} \\
$K^{*}(1410)$ & $C^{*}_{2}$ &  $1^{-}$ &$\frac{1}{2}$&
 $2.003$ & $0.175\pm 0.014$ &
$4.1\pm 0.6$ & $1.14\pm 0.18$ \\
$K^{*}_{0}(1950)$ & $E_{2}$ &  $0^{+}$&$\frac{1}{2}$ &
 \multicolumn{4}{c|}{forbidden} \\
$K_{1}(1650)$ & $E^{*}_{2}$ &  $1^{+}$ &$\frac{1}{2}$&
 $1.756$ & $0.229\pm 0.040$ &
$1.7\pm 0.6$ & $0.47\pm 0.16$ \\
\multicolumn{6}{|c}{total} &$37.4\pm 13.6$ & $10.44\pm 3.78$\\
\hline
\end{tabular}
\end{center}
\end{table}

\begin{table}
\caption{Comparison of our results for the ratio
$R=\frac{\Gamma(B\rar K^{**}\gamma)}{\Gamma(B\rar X_{s}\gamma)}$
with the previous work done in
\protect\cite{mannel} and \protect\cite{altomari}. Note that
in the quark model calculations decay into the $^{1}P_{1}$ state
is forbidden, because ${\cal O}_{7}$ is a spin-flip operator, and
$K_{1}(1270)$ and $K_{1}(1400)$ are mixtures of $^{1}P_{1}$ and
$^{3}P_{1}$ states. In \protect\cite{altomari} $^{3}P_{1}$ state
had $R=6\%$.}
\label{tab2}
\begin{center}
\begin{tabular}{|cccccc|}
\hline
Meson         &     State     &   $J^{P}$  &   $R[\%]$ (this work) &
$R[\%]$ (ref. \protect\cite{mannel}) & $R[\%]$
 (ref. \protect\cite{altomari})    \\
\hline
$K$    &  $C$     & $0^{-}$  & \multicolumn{3}{c|}{forbidden} \\
$K^{*}(892)$ & $C^{*}$ &  $1^{-}$ &
$16.8\pm 6.4$ & $3.5-12.2$ & 4.5\\
$K^{*}(1430)$ & $E$ &  $0^{+}$&
 \multicolumn{3}{c|}{forbidden} \\
$K_{1}(1270)$ & $E^{*}$ &  $1^{+}$ &
$4.3\pm 1.6$ & $4.5-10.1$ & forbidden/6.0 \\
$K_{1}(1400)$ & $F$ &  $1^{+}$ &
$2.1\pm 0.9$ & $6.0-13.0$ &forbidden/6.0\\
$K^{*}_{2}(1430)$ & $F^{*}$ &  $2^{+}$ &
$6.2\pm 2.9$ & $17.3-37.1$ & 6.0\\
$K^{*}(1680)$ & $G$ &  $1^{-}$ &
$0.5\pm 0.2$ & $1.0-1.5$ & 0.9\\
$K_{2}(1580)$ & $G^{*}$ &  $2^{-}$ &
$1.7\pm 0.4$ & $4.5-6.4$ & 4.4 \\
$K(1460)$    &  $C_{2}$     & $0^{-}$  &
\multicolumn{3}{c|}{forbidden} \\
$K^{*}(1410)$ & $C^{*}_{2}$ &  $1^{-}$ &
$4.1\pm 0.6$ & $7.2-10.6$ & 7.3\\
$K^{*}_{0}(1950)$ & $E_{2}$ &  $0^{+}$ &
\multicolumn{3}{c|}{forbidden} \\
$K_{1}(1650)$ & $E^{*}_{2}$ &  $1^{+}$ & $1.7\pm 0.6$ & not given &
not given \\
\multicolumn{3}{|c}{total} & $37.4\pm 13.6 $&
$44.1 - 90.9$ & $29.1$ \\
\hline
\end{tabular}
\end{center}
\end{table}

\begin{table}
\caption{Comparison of our results for the ratio
$R=\frac{\Gamma(B\rar K^{*}\gamma)}{\Gamma(B\rar X_{s}\gamma)}$
with several previous calculations.}
\label{tab3}
\begin{center}
\begin{tabular}{|lcc|lcc|}
\hline
Author(s) & Ref. & $R[\%]$ \\
O'Donnell (1986) & \protect\cite{donnell} & 97.0
\\
Deshpande et al. (1988) &\protect\cite{deshpande2} & 6.0
\\
Dominguez et al. (1988) &\protect\cite{dominguez} & $28.0\pm 11.0$
 \\
Altomari (1988) &\protect\cite{altomari} & 4.5
 \\
Deshpande et al. (1989) &\protect\cite{deshpande3} & 6.0-14.0
\\
Aliev et al. (1990) &\protect\cite{aliev} & 39.0
 \\
Ali et al. (1991) &\protect\cite{ali2} & 28.0-40.0
\\
Du et al. (1992) &\protect\cite{du} & 69.0
 \\
Faustov et al. (1992) &\protect\cite{faustov} & 6.5
\\
El-Hassan et al. (1992) &\protect\cite{hassan} & 0.7-12.0
\\
O'Donnell et al. (1993) & \protect\cite{donnell2} & 10.0
\\
Colangelo et al. (1993) & \protect\cite{colangelo} & $17.0\pm 5.0$
\\
 Ali et al. (1993) &\protect\cite{mannel} & $3.5\pm 12.2$
\\
Ali et al. (1993) &\protect\cite{simma} & $16.0\pm 5.0$
\\
Ali et al. (1993) &\protect\cite{ali93} & $13.0\pm 3.0$
\\
 Ball (1994) &\protect\cite{ball} & $20.0\pm 6.0$
\\
Narison (1994) &\protect\cite{narison} & $16.0\pm 4.0$
\\
Holdom et al. (1994) &\protect\cite{holdom} & $17.0\pm 4.0$
\\
Atwood et al. (1994) &\protect\cite{atwood} & 1.6-2.5
\\
Bernard et al. (1994) &\protect\cite{bernard} & $6.0\pm1.2\pm 3.4$
\\
Ciuchini et al. (1994) &\protect\cite{ciuchini} & $23.0\pm9.0$
\\
Bowler et al. (1994) &\protect\cite{bowler} & $9.0\pm 3.0\pm 1.0$
\\
Burford et al. (1995) &\protect\cite{burford} & 15.0-35.0
\\
Tang et al. (1995) &\protect\cite{tang} & 10.0-12.0 \\
\multicolumn{2}{|c}{this work} & $16.8\pm 6.4$ \\
\hline
\end{tabular}
\end{center}
\end{table}

\clearpage

\newpage
\begin{figure}
\begin{center}
FIGURES
\end{center}
\vskip 1cm
\caption{The experimental inclusive $B\rar X_{s}\gamma$
mass distribution measured at $CLEO$ \protect\cite{alam}.
The data  have
been normalized to unity. The curve is the sum of the exclusive
$K^{**}\gamma$ channels from Table \protect\ref{tab1} as calculated
by (\protect\ref{md}).}
\label{fig1}
\end{figure}

\newpage

\begin{figure}[p]
\epsfxsize = 5.4in \epsfbox{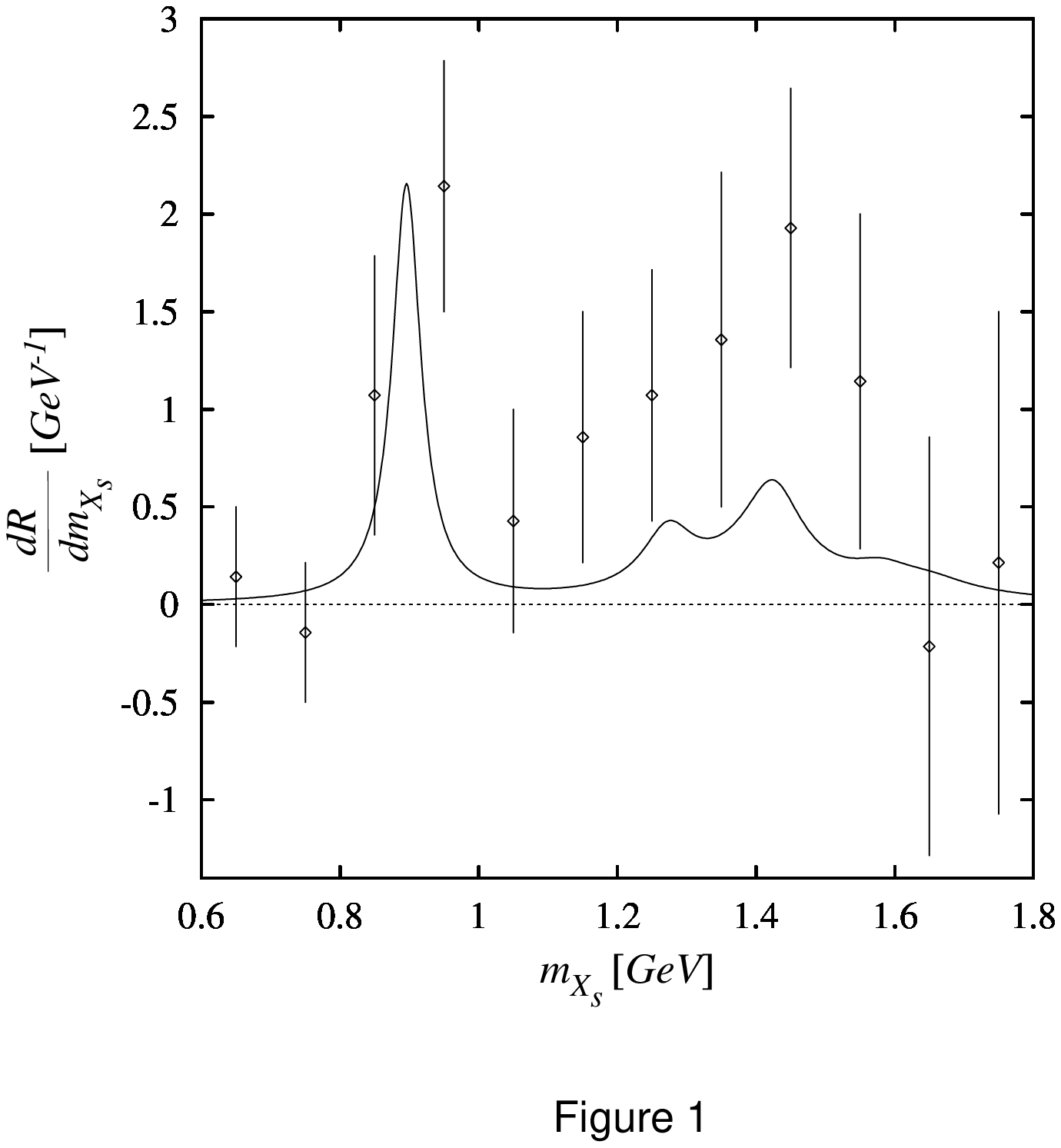}
\end{figure}


\newpage

\begin{thebibliography}{99}

\bibitem{shifman} A. I Vainshtein, V. I Zakharov, and M. A. Shifman,
JETP Lett. {\bf 22}, 55 (1975);
J. Ellis, M. K. Gaillard, and D. V. Nanopoulos,
Nucl. Phys. B {\bf 100}, 313 (1975);
M. Bander, D. Silverman, and A. Soni,
Phys. Rev. Lett. {\bf 43}, 242 (1979).
\bibitem{hewett} J. Hewett, {\em Top Ten Models
Constrained by $b\rar s\gamma$}, preprint
SLAC-PUB-6521 (hep-ph/9406302).
\bibitem{ammar} R. Ammar et al., CLEO Collaboration,
Phys. Rev. Lett. {\bf 71}, 674 (1993).
\bibitem{cleoex} K. W. Edwards et. al.,
CLEO Colaboration, {\em Update on  Exclusive
$B^{0}\rar K^{*0}(892)\gamma $ Measurement at CLEO},
preprint CLEO CONF95-6, July 1995.
\bibitem{alam} M. S. Alam et al., CLEO Collaboration
Phys. Rev. Lett. {\bf 74}, 2885 (1995).
\bibitem{ali2} A. Ali and T. Mannel,
Phys. Lett. B {\bf 264}, 447 (1991);
Phys. Lett. B {\bf 274}, 526(E) (1992).
\bibitem{mannel} A. Ali, T. Ohl, and T. Mannel,
Phys. Lett. B {\bf 298}, 195 (1993).
\bibitem{dominguez} C. A. Dominguez, N. Paver, and Riazuddin,
Phys. Lett. B {\bf 214}, 459 (1988).
\bibitem{aliev} T. M. Aliev, A. A. Ovchinnikov, and  V. A. Slobodenyuk,
Phys. Lett. B {\bf 237}, 569 (1990).
\bibitem{colangelo} P. Colangelo, C.A. Dominguez, G. Nardulli, and
 N. Paver, Phys. Lett. B {\bf 317}, (1993).
\bibitem{ball} P. Ball, {\em The Decay
$B\rar K^{*}\gamma$ from QCD Sum Rules}, preprint TUM-T31-43/93
(hep-ph/9308244).
\bibitem{simma} A. Ali, V. M. Braun, and H. Simma,
Z. Phys. C {\bf 63}, 437 (1994).
\bibitem{narison} S. Narison, Phys. Lett. B {\bf 327},
354 (1994).
\bibitem{donnell} P. J. O'Donnell, Phys. Lett. B {\bf 175}, 369 (1986).
\bibitem{deshpande2} N. G. Deshpande, P. Lo, J. Trampeti\'{c},
 G. Eilam,  and P. Singer,
Phys. Rev. Lett. {\bf 59}, 183 (1987).
\bibitem{altomari} T. Altomari,
Phys. Rev. D {\bf 37}, 677 (1988).
\bibitem{donnell2} P. J. O'Donnell and H. K. Tung,
Phy. Rev. D {\bf 48}, 2145 (1993).
\bibitem{ali93} A. Ali and C. Greub,
Z. Phys. C {\bf 60}, 433 (1993).
\bibitem{deshpande3} N. G. Deshpande and J. Trampeti\'{c},
Mod. Phys.  Lett. A {\bf 4}, 2095 (1989).
\bibitem{du} D. Du and C. Liu, {\em The $1/m_{s}$
 Corrections to the  Exclusive Rare $B$ Meson  Decays},
preprint BIHEP-TH-92-41.
\bibitem{faustov} R. N. Faustov and V. O. Galkin,
Mod. Phys. Lett. A {\bf 7}, 2111 (1992).
\bibitem{hassan} E. El-Hassan and Riazuddin,
Phys. Rev. D {\bf 47}, 1026 (1993).
\bibitem{holdom} B. Holdom and  M. Sutherland,
Phys. Rev. D {\bf 49}, 2356 (1994).
\bibitem{tang} J. Tang, J. Liu, and K. Chao,
 Phys. Rev. D {\bf 51}, 3501 (1995).
\bibitem{atwood} D. Atwood and A. Soni,
Z. Phys. C {\bf 64}, 241 (1994).
\bibitem{bernard} C. Bernard, P. Hsieh, and A. Soni,
Phys. Rev. Lett. {\bf 72}, 1402 (1994).
\bibitem{ciuchini} M. Ciuchini, E. Franco, G. Martinelli, L. Reina, and
L. Silvestrini,
Phy. Lett. B {\bf 334}, 137 (1994).
\bibitem{bowler} K.C. Bowler et al., {\em A
 lattice calculation of the  branching ratio
for some of the exclusive modes of $b\rar s\gamma$},
preprint EDINBURGH-94-544 (hep-lat/9407013).
\bibitem{burford} D. R. Burford et al., {\em
Form-factors for $B\rar \pi l\bar{\nu}_{l}$ and
$B\rar K^{*}\gamma $ decays on the lattice},
preprint
FERMILAB-PUB-95/023-T (hep-lat/9503002).
\bibitem{georgi2} H. Georgi, Nucl. Phys. B {\bf 348}, 293 (1991).
\bibitem{korner} J. G. K$\ddot{\rm o}$rner and G. A. Schuler,
Z. Phys. C {\bf 38}, 511 (1988).
\bibitem{falk} A. F. Falk,
Phys. Lett. B {\bf 378}, 79 (1992).
\bibitem{modelling} S. Veseli and M. G. Olsson, {\em
Modelling form factors in HQET}, UW-Madison preprint MADPH-95-899
(hep-ph/9507425).
\bibitem{grinstein} B. Grinstein, R. Springer, and M. B. Wise,
Nucl. Phys. B {\bf 339}, 269 (1990).
\bibitem{ali} A. Ali and C. Greub,
Z. Phys. C {\bf 49}, 431 (1991).
\bibitem{buras} A. J. Buras, M. Misiak, M. M$\ddot{\rm u}$nz, and
S. Pokorski,
Nucl. Phys. B {\bf 424}, 374 (1994).
\bibitem{deshpande} N. G. Deshpande, {\em Theory of Penguins
in B Decay}, in {\em B Decays}, 2nd edition, edt. by S. Stone,
World Scientific, 1994.
\bibitem{zalewski2} M. Sadzikowski and K. Zalewski,
Z. Phys. C {\bf 59}, 677 (1993).
\bibitem{isgw} N. Isgur, D. Scora, B. Grinstein, and
M. B. Wise, Phys. Rev. D {\bf 39}, 799 (1989).
\bibitem{cleo} B. Barish et al., CLEO Collaboration,
Phys. Rev. D {\bf 51}, 1014 (1995).
\bibitem{duncan} A. Duncan, E. Eichten, and H. Thacker,
Phys. Lett. B {\bf 303}, 109 (1993).
\bibitem{iw} M. G. Olsson and S. Veseli,
 Phys. Lett. B. {\bf 353}, 96 (1995).
\bibitem{pdg} Particle Data Group, Phys. Rev. D {\bf 50},
1173 (1994).
\end{thebibliography}
\end{document}